# Ultra-low damping in lift-off structured yttrium iron garnet thin films

A. Krysztofik,[1] L. E. Coy,[2] P. Kuświk,[1,3] K. Załęski,[2] H. Głowiński,[1] and J. Dubowik[1]

[1]*Institute of Molecular Physics, Polish Academy of Sciences, PL-60-179 Poznań, Poland*
[2]*NanoBioMedical Centre, Adam Mickiewicz University, PL-61-614 Poznań, Poland*
[3]*Centre for Advanced Technology, Adam Mickiewicz University, PL-61-614 Poznań, Poland*
Electronic mail: adam.krysztofik@ifmpan.poznan.pl, hubert.glowinski@ifmpan.poznan.pl

We show that using maskless photolithography and the lift-off technique patterned yttrium iron garnet thin films possessing ultra-low Gilbert damping can be accomplished. The films of the 70 nm thickness were grown on (001)-oriented gadolinium gallium garnet by means of pulsed laser deposition and exhibit high crystalline quality, low surface roughness and effective magnetization of 127 emu/cm$^3$. The Gilbert damping parameter is as low as $5 \times 10^{-4}$. The obtained structures have well-defined sharp edges which along with good structural and magnetic film properties, pave a path in the fabrication of high-quality magnonic circuits as well as oxide-based spintronic devices.

Yttrium iron garnet (Y$_3$Fe$_5$O$_{12}$, YIG) has become an intensively studied material in recent years due to exceptionally low damping of magnetization precession and electrical insulation enabling its application in research on spin-wave propagation[1–3], spin-wave based logic devices[4–6], spin pumping[7], and thermally-driven spin caloritronics[8]. These applications inevitably entail film structurization in order to construct complex integrated devices. However, the fabrication of high-quality thin YIG films requires deposition temperatures over 500°C[6,9–18] leading to top-down lithographical approach that is ion-beam etching of a previously deposited plain film whereas patterned resist layer serves as a mask. Consequently, this method introduces crystallographic defects, imperfections to surface structure and, in the case of YIG films, causes significant increase of the damping parameter.[19–21] Moreover, it does not ensure well-defined structure edges for insulators, which play a crucial role in devices utilizing edge spin waves[22], Goos-Hänchen spin wave shifts[23,24] or standing spin waves modes[25]. On the contrary, the bottom-up structurization deals with these issues since it allows for the film growth in the selected, patterned areas followed by a removal of the resist layer along with redundant film during lift-off process. Additionally, it reduces the patterning procedure by one step, that is ion etching, and imposes room-temperature deposition which both are particularly important whenever low fabrication budget is required.

In this letter we report on ultra-low damping in the bottom-up structured YIG film by means of direct writing photolithography technique. In our case, the method allows for structure patterning with 0.6 μm resolution across full writing area. In order to not preclude the lift-off process, the pulsed laser deposition (PLD) was conducted at room temperature and since such as-deposited films are amorphous[19,27] the *ex-situ* annealing was performed for recrystallization. Note that post-deposition annealing of YIG films is commonly carried out regardless the substrate temperature during film deposition[6,12,13,28,29]. As a reference we investigated a plain film which was grown in the same deposition process and underwent the same fabrication procedure except for patterning. Henceforth, we will refer to the structured and the plain film as Sample 1 and Sample 2, respectively. We anticipate that such a procedure may be of potential for fabrication of other magnetic oxide structures useful in spintronics.

Structural characterization of both samples was performed by means of X-Ray Diffraction (XRD). Atomic force microscopy (AFM) was applied to investigate surface morphology and the quality of structure edges. SQUID magnetometry provided information on the saturation magnetization and magnetocrystalline anisotropy field. Using a coplanar waveguide connected to a vector network analyzer, broadband ferromagnetic resonance (VNA-FMR) was performed to determine Gilbert damping parameter and anisotropy fields. All the experiments were conducted at the room temperature.

The procedure of samples preparation was as follows. The (001)-oriented gadolinium gallium garnet substrates were ultrasonicated in acetone, trichloroethylene and isopropanol to remove surface impurities. After a 1 minute of hot plate baking for water evaporation, a positive photoresist was spin-coated onto the substrate (Sample 1). Using maskless photolithography an array of 500 μm x 500 μm squares separated over 500 μm was patterned and the exposed areas were developed. Detailed parameters of photolithography process can be found in Ref.[26]. We chose rather large size of the squares to provide a high signal-to-noise ratio in the latter measurements. Thereafter, plasma etching was performed to remove a residual resist. We would like to emphasize the importance of this step in the fabrication procedure as the resist residues may locally affect crystalline structure of a YIG film causing an undesirable increase of overall magnetization damping. Both substrates were then placed in a high vacuum chamber of $9\times10^{-8}$ mbar base pressure and a film was deposited from a stoichiometric ceramic YIG target under $2\times10^{-4}$ mbar partial pressure of oxygen. We used a Nd:YAG laser (λ = 355 nm) for the ablation with pulse rate of 2 Hz which yielded 1 nm/min growth rate. The target-to-substrate distance was approximately 50 mm. After the deposition the lift-off process for the Sample 1 was performed using sonication in acetone to obtain the expected structures. Subsequently, both samples were annealed in a tube furnace under oxygen atmosphere (p ≈ 1 bar) for 30 minutes at 850°C. The heating and cooling rates were about 50 °C/min and 10 °C/min, respectively.

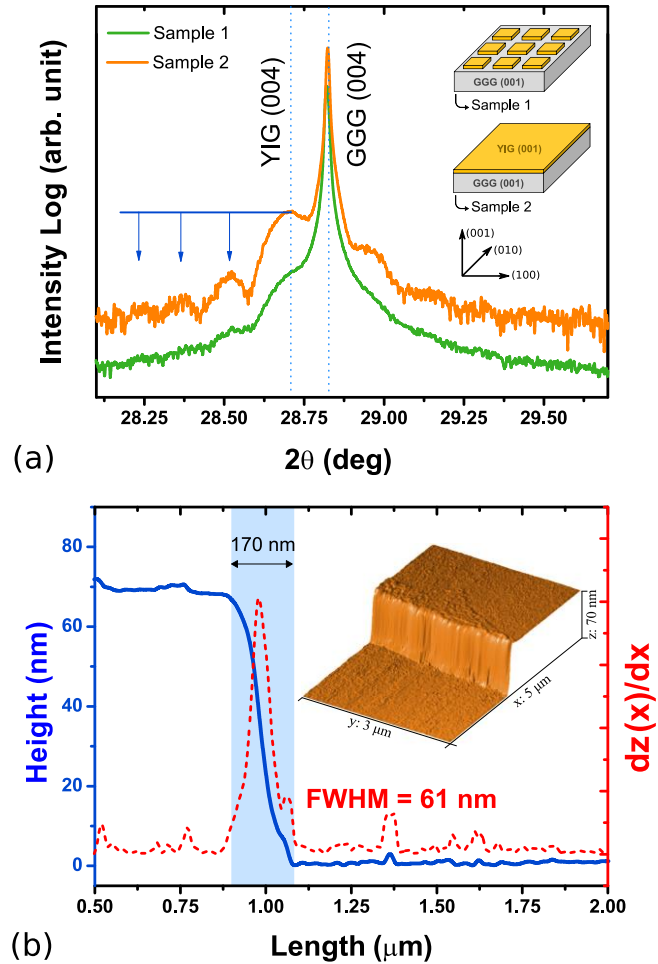

FIG. 1. (a) XRD θ−2θ plot near the (004) reflection of structured (Sample 1) and plain (Sample 2) YIG film. Blue arrows show clear Laue reflections of the plain film. Insets show schematic illustration of the structured and plain film used in this study. (b) Height profile (z(x)) taken from the structured sample (left axis), right shows the differential of the profile, clearly showing the slope change. Inset shows 3D map of the structure's edge.

The structure of YIG films was determined by the X-ray diffraction. Although the as-deposited films were amorphous, with the annealing treatment they inherited the lattice orientation of the GGG substrate and recrystallized along [001] direction. Figure 1 (a) presents diffraction curves taken in the vicinity of (004) Bragg reflection. The (004) reflection position of structured YIG well coincides with the reflection of the plain film. The 2θ=28.709° corresponds to the cubic lattice constant of 12.428 Å. A comparison of this value with lattice parameter of a bulk YIG (12.376 Å) suggest distortion of unit cells due to slight nonstoichiometry.[16,30] Both samples exhibit distinct Laue oscillations depicted by the blue arrows, indicating film uniformity and high crystalline order, although the structured film showed lower intensity due to the lower mass of the film. From the oscillation period we estimated film thickness of 73 nm in agreement with the nominal thickness and the value determined using AFM for Sample 1 (Fig. 1 (b)). By measuring the diffraction in the expanded angle range we also confirmed that no additional phases like $Y_2O_3$ or $Fe_2O_3$ appeared.

The surface morphology of the structured film was investigated by means of AFM. In Fig. 1 (b) profile of a square's edge is shown. It should be highlighted that no edge irregularities has formed during lift-off process. The horizontal distance between GGG substrate and the surface of YIG film is equal to 170 nm as marked in Fig. 1 (b) by the shaded area. A fitting with Gaussian function to the derivative of height profile yields the full width at half maximum of 61 nm. This points to the well-defined structure edges achieved with bottom-up structurization. Both samples have smooth and uniform surfaces. The comparable values of root mean square (RMS) roughness (0.306 nm for Sample 1 and 0.310 nm for Sample 2) indicate that bottom-up structurization process did not leave any resist residues. Note that a roughness of a bare GGG substrate before deposition was 0.281 nm, therefore, the surface roughness of YIG is increased merely by 10%.

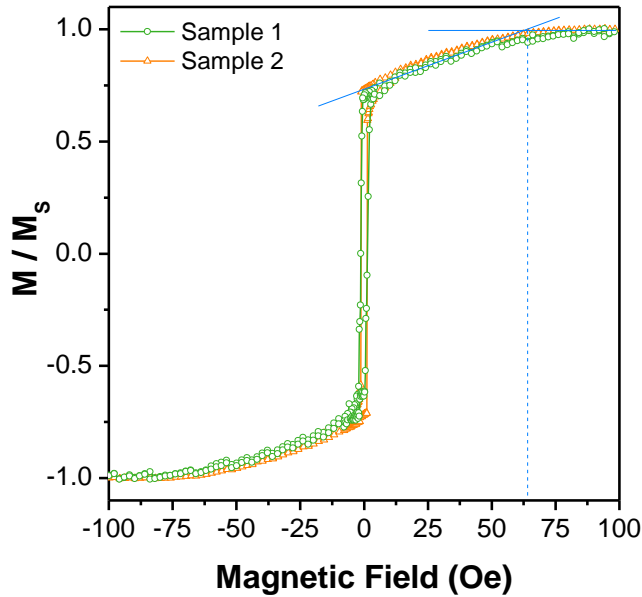

FIG. 2. Hysteresis loops of structured (Sample 1) and plain (Sample 2) YIG films measured by SQUID magnetometry along [100] direction at the room temperature.

Figure 2 shows magnetization reversal curves measured along [100] direction. For each hysteresis loop a paramagnetic contribution arising for the GGG substrates was subtracted. The saturation magnetization $M_s$ was equal to 117 emu/cm$^3$ and 118.5 emu/cm$^3$ for Sample 1 and 2, respectively. Both hysteresis loops demonstrate in-plane anisotropy. For the (001)-oriented YIG the [100] direction is a "hard" in-plane axis and the magnetization saturates at $H_a = 65$ Oe. This value we identify as magnetocrystalline anisotropy field. The VNA-FMR measurements shown in Fig. 3 (a) confirm these results. Using Kittel dispersion relation, i.e. frequency $f$ dependence of resonance magnetic field $H$:

$$f = \frac{\gamma}{2\pi}\sqrt{(H + H_a \cos 4\varphi)\left(H + \frac{1}{4}H_a(3 + \cos 4\varphi) + 4\pi M_{eff}\right)}, \qquad (1)$$

$$4\pi M_{eff} = 4\pi M_s - H_u, \qquad (2)$$

we derived $H_a$ and the effective magnetization $M_{eff}$, both comparable to the values determined using SQUID and close to the values of a bulk YIG (see Table I.). Here, the azimuthal angle $\varphi$ defines the in-plane orientation of the magnetization direction with respect to the [100] axis of YIG and $\gamma$ is the gyromagnetic ratio ($1.77 \times 10^7 G^{-1}s^{-1}$). To better compare the values of $H_a$ between samples and to determine if the results are influenced by additional anisotropic contribution arising from the squares' shape in the structured film we performed angular resolved resonance measurements (inset in Fig. 3(a)). The fitting according to Eq. (1) gives $|H_a|$ equal to 69.5±0.6 for Sample 1 and 69.74±0.28 for Sample 2 in agreement with the values derived from $f(H)$ dependence and better accuracy. Hence, we conclude that the structurization did not affect the in-plane anisotropy. The deviations of the derived $M_s$ and $H_a$ from bulk values can be explained in the framework of Fe vacancy model developed for YIG films as a result of nonstoichiometry.[13,30] For the experimentally determined $M_s$ and $H_a$ the model yields the chemical unit $Y_3Fe_{4.6}O_{11.4}$ which closely approximates to the composition of a stoichiometric YIG $Y_3Fe_5O_{12}$.

TABLE I. Key parameters reported for PLD and LPE YIG films.

| | | AFM | SQUID | | VNA-FMR | | | | | |
|---|---|---|---|---|---|---|---|---|---|---|
| | Film thickness | RMS roughness (nm) | $M_s$ (emu/cm$^3$) | $H_a$ (Oe) | Field orientation | $M_{eff}$ (emu/cm$^3$) | $|H_a|$ (Oe) | $H_u$ (Oe) | $\alpha$ ($\times 10^{-4}$) | $\Delta H_0$ (Oe) |
| Sample 1 | 70 nm | 0.306 | 117±1 | 65±5 | (100): | 125±1 | 64±1 | -101±18 | 5.53±0.13 | 1.45±0.09 |
| | | | | | (110): | 126±1 | 63±1 | -113±18 | 5.24±0.12 | 2.86±0.09 |
| | | | | | (001): | 129±2 | − | -151±28 | 5.19±0.64 | 2.61±0.34 |
| Sample 2 | 70 nm | 0.310 | 118.5±2 | 65±5 | (100): | 124±1 | 62±1 | -69±28 | 5.05±0.07 | 0.97±0.05 |
| | | | | | (110): | 127±1 | 65±1 | -107±28 | 5.09±0.09 | 1.28±0.06 |
| | | | | | (001): | 131±2 | − | -157±36 | 5.02±0.18 | 1.48±0.09 |
| LPE-YIG[31] | 106 nm | 0.3 | 143 | − | (112): | − | − | − | 1.2 | 0.75 |
| LPE-YIG[30] | 120 μm | − | 139±2 | − | (111): | 133±2 | 85±6 | 76±1 | 0.3 | − |

Although the saturation magnetization of the films is decreased by 15% with respect to the bulk value we can expect similar spin wave dynamics since magnon propagation does not solely depend on $M_s$ but on the effective magnetization or equivalently, on the uniaxial anisotropy field $H_u$.[12] Substitution of $M_s$ into Eq. (2) gives average values of $H_u$ equal to -122 Oe and -111 Oe for Sample 1 and 2, respectively (to determine $H_u$ from the out-of-plane FMR measurements when H ∥ [001] we used the $f = \frac{\gamma}{2\pi}(H + H_a - 4\pi M_{eff})$ dependence[13] to fit the data and assumed the value of $H_a$ from angular measurements). As $M_{eff}^{Sample\,1,2} \approx M_{eff}^{bulk}$, it follows that the low value of $M_s$ in room-temperature deposited thin films is "compensated" by uniaxial anisotropy field. Note that for bulk YIG saturation magnetization is diminished by $H_u/4\pi$ giving a lower value of $M_{eff}$ while for Sample 1 and 2, $M_s$ is augmented by $H_u/4\pi$ giving a higher value of $M_{eff}$ (Table I.). The negative sign of uniaxial anisotropy field is typical for PLD-grown YIG films and originates from preferential distribution of Fe vacancies between different sites of YIG's octahedral sublattice.[30] This points to the growth-induced anisotropy mechanism while the stress-induced contribution is of ≈10 Oe[29] and, as it can be estimated according to Ref.[32], the transition layer at the substrate-film interface due to Gd, Ga, Y ions diffusion is ca. 1.5 nm thick for the 30 min of annealing treatment. We argue that the growth-induced anisotropy due to ordering of the magnetic ions is related to the growth condition which in our study is specific. Namely, it is crystallization of an amorphous material.

Gilbert damping parameter $\alpha$ was obtained by fitting dependence of linewidth $\Delta H$ (full width at half maximum) on frequency $f$ as shown in Fig. 3 (b):

$$\Delta H = \frac{4\pi\alpha}{\gamma}f + \Delta H_0, \qquad (3)$$

where $\Delta H_0$ is a zero-frequency linewidth broadening. The $\alpha$ parameter of both samples is nearly the same, $5.32 \times 10^{-4}$ for Sample 1 and $5.05 \times 10^{-4}$ for Sample 2 on average (see Table I.). It proves that bottom-up patterning does not compromise magnetization damping. The value of $\Delta H_0$ contribution is around 1.5 Oe although small variations of $\Delta H_0$ on $\varphi$ can be noticed. Additional comments on angular dependencies of $\Delta H$ can be found in the supplementary material. The derived values of $\alpha$ remain one order of magnitude smaller than for soft ferromagnets like $Ni_{80}Fe_{20}$[33], CoFeB[34] or Finemet[35], and are comparable to values reported for YIG films deposited at high temperatures (from $1 \times 10^{-4}$ up to $9 \times 10^{-4}$).[6,9,11,14,15,17,18] It should be also highlighted that $\alpha$ constant is significantly increased in comparison to the bulk YIG made by means of Liquid Phase Epitaxy (LPE). However, recently reported LPE-YIG films of nanometer thickness, suffer from the increased damping as well (Table I.) due to impurity elements present in the high-temperature solutions used in LPE technique[31]. As PLD method allows for a good contamination control, we attribute the increase as a result of slight nonstoichiometry determined above with Fe vacancy model.[30] Optimization of growth conditions, which further improve the film composition may resolve this issue and allow to cross the $\alpha = 1 \times 10^{-4}$ limit. We also report that additional annealing of the samples (for 2h) did not influence damping nor it improved the value of $H_a$ or $M_{eff}$ (within 5% accuracy).

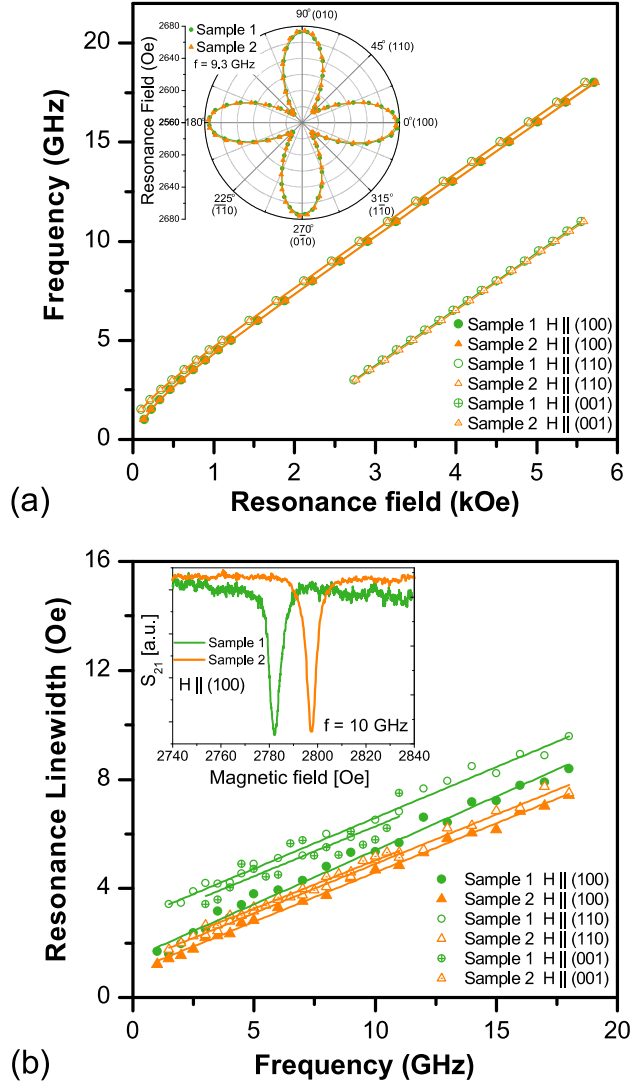

FIG. 3. (a) Kittel dispersion relations of the structured (Sample 1) and plain (Sample 2) YIG film. The inset shows angular dependence of resonance field revealing perfect fourfold anisotropy for both samples. (b) Linewidth dependence on frequency fitted with Eq. (3). The inset shows resonance absorptions peaks with very similar width (5.3 Oe for Sample 1 and 4.7 Oe for Sample 2 at 10 GHz). Small differences of the resonance field originate from different values of $4\pi M_{eff}$.

In conclusion, the lift-off patterned YIG films possessing low damping have been presented. Although the structurization procedure required deposition at room temperature, the $\alpha$ parameter does not diverge from those reported for YIG thin films grown at temperatures above 500°C. Using the plain, reference film fabricated along with the structured one, we have shown that structurization does not significantly affect structural nor magnetic properties of the films, i.e. out-of-plane lattice constant, surface roughness, saturation magnetization, anisotropy fields and damping. The structures obtained with bottom-up structurization indeed possess sharp, well-defined edges. In particular, our findings will help in the development of magnonic and spintronic devices utilizing film boundary effects and low damping of magnetization precession.

Supplementary Material

See supplementary material for the angular dependence of resonance linewidth.

The research received funding from the European Union Horizon 2020 research and innovation programme under the Marie Skłodowska-Curie grant agreement No 644348 (MagIC). We would like to thank Andrzej Musiał for the assistance during film annealing.